# An approach to define Very High Capacity Networks with improved quality at an affordable cost

**Giovanni Santella**
Autorità per le garanzie nelle comunicazioni
Napoli, ITALY

**Francesco Vatalaro**
Dipartimento di Ingegneria dell'Impresa
"Mario Lucertini"
Università di Roma Tor Vergata
Roma, ITALY

*Abstract*—This paper introduces an approach to identify meaningful VHCNs (Very High Capacity Networks) network KPIs (Key Performance Indicators) to promote efficient operator investments and improve the end-user Quality of Experience (QoE). By selecting appropriate design parameters, we can achieve this critical objective while reducing, at the same time, the risk of inefficient investments. In the paper, we clarify that, especially for VHCN networks, what it is most relevant to customer satisfaction is the throughput at the application layer (generically indicated as "bandwidth") much more than the data-rate at the physical or IP layer. The reason for this is that bandwidth is strictly linked to latency. Consequently, the most demanding services envisaged for the future (e.g., mobile virtual and augmented reality, tactile internet) cannot be provided by merely increasing data-rate. As a mandatory pre-requisite, they need overall appropriate network design to reduce latency, as an example, through Edge Cloud Computing (ECC).

## I. Introduction

The objective of 1 Gbps bandwidth, and even more, is one necessary target for some of the challenging services envisaged within the present decade (e.g., mobile virtual and augmented reality, tactile internet, connected autonomous vehicle systems). On the one hand, they are not for tomorrow and, luckily enough, we still have some time to reach the target anywhere and anytime. On the other hand – and this is a crucial concept to be made clear as soon as possible – we must interpret the bandwidth as throughput at the application layer rather than bit-rate at the physical (or network) layer. The consequence is twofold. The objective is even more challenging than expected as it is not just a matter of laying more optical fiber to increase "speed." Simultaneously, we should use the timeframe remaining for the consolidation of new services with gradual investment planning to match the technological evolution and the increased demand, aiming at the vital objective of not leaving any citizens behind.

In the following, some basic concepts are clarified regarding:

- Quality of Experience (QoE) for the customer and the difference with traditional Quality of Service (QoS) concept
- KPIs needed to evaluate the level of customer satisfaction, and
- The criticality stemming from the dependence of throughput on latency.

This paper is concluded with some suggestions deemed useful for Regulators and, in the EU, for the European Commission and the BEREC (Body of European Regulators for Electronic Communications) to consider in the framework of the guidelines on VHCNs (Very High Capacity Networks). Main objective here is to suggest how the VHCNs implementation can be carried out, so speeding up the match of the future vision of communications networks for Europe and elsewhere.

## II. The Urgent Need to Invest in QoE

The objective of a Telco operator is customer satisfaction. This obvious remark brings about the concept of QoE. In turn, although it is generally well-known, this concept sometimes may tend to be confused with that of QoS. Let us try to clarify the matter.

### A. QoE and QoS

A rather broad definition of Quality of Experience, provided by ITU-T P.10-2016, consists of "the degree of delight or annoyance of the user of an application or service. It results from the fulfillment of his or her expectations with respect to the utility and/or enjoyment of the application or service in the light of the user's personality and current state." Although not always numerically quantifiable, QoE is the most significant factor in assessing customer experience.

According to ITU-T P.10/G.100, QoS is the "totality of characteristics of a telecommunications service that bear on its ability to satisfy stated and implied needs of the user of the service."

When compared, the two definitions show a substantial difference. On the one hand, QoS is network-centric and provides an "ensemble view" on totality of characteristics – i.e., it is based on average values – not directly related to the service experience a given customer perceives. Traditionally, its goal is to support network management and improve the average network quality. In general, QoS cannot appropriately meet the expectations of one specific customer, who demand peak-hour rather than average time quality. On the other hand, QoE focuses on performance (or quality) of applications, both at technical and subjective level. As it addresses customer's view, it is not appropriate for network maintenance purposes, the task of QoS. Therefore, both QoE and QoS are important in network design, and their roles are not interchangeable but complementary.

QoE provides a subjective (but not arbitrary) set of measures from the user's point of view of the overall quality of the service provided, aiming at capturing specific needs. Therefore, QoE is user-centric. In other words, rather than focusing only on technical parameters, QoE provides both an overall and detailed assessment of human being expectations, feelings, perceptions, cognition, and satisfaction with any particular application or service. Customer's satisfaction is related both to technical and nontechnical issues [1]. We evaluate technical QoE through so-called applications KPIs. However, network KPIs are not univocally associated with them. Therefore, the traditional Telcos attempts to infer QoE by linking it to QoS generally provide poor results.

When considering KPIs suitable to assess technical QoE, as their values are closely related to the user's satisfaction, we intend to evaluate them in the peak-hour. Moreover, they must be assessed for each customer one by one.

When technical parameters are averaged over several users, they are QoS parameters. They are instrumental in supporting network management to supervise network availability. However, they are generally of little use to ensure customer satisfaction.

As a first catch-up, it is evident that it is worth investing in QoE in order to favor efficient expenditure for a Telco and improve customer satisfaction, while at the same time increasing financial stability of the telecoms ecosystem as a whole. Investments on QoS are certainly needed, but they are a cost for the Telco's network management; contrarily, QoE pave the way to new business models and improve infrastructures monetization.

*B. Services and key performance indicators*

Telecommunications networks provide two types of services, application services (or, more simply, applications) and network services:

1. *Application services* are the end-users' applications. Typically, to provide applications a software runs on the servers of service providers and devices used by end-users themselves. Today, mostly OTTs offer application services and, in a limited way, other Content Delivery Providers (CDPs).
2. *Network services* are the IP packet transport services that a Telco operator must provide to achieve the performance levels required by applications.

Among applications, video services are the most important ones to monitor today, as they are the most demanding and those requiring the highest fraction of transmission resources. Video streaming over the internet requires a share of bandwidth expected to grow over 82% of all consumer internet traffic in 2022 (Source: Cisco VNI).

We classify key performance indicators as applications KPIs and network KPIs.

Applications KPIs are related to the performance of all the applications and, in particular, of the video services. For video services, applications KPIs are [1]:

- *Throughput*, the "speed" at which end-user devices and servers exchange application data. It is one of the most important indicators and can reach up to 1 Gbps. Throughput is always lower than the bit-rate (BR), the "speed" in the communications channel between end-user and server, due to the congestion control algorithms. For current UBB (Ultra Broadband) networks, this difference is huge (Throughput << BR). The BR and – when the BR is not the bottleneck – the 'distance' metric between the application and the end-user limit the throughput (see below, eq. (2)).
- *Latency*, measured through the round-trip time (RTT). Latency must be very low (i.e., one millisecond, or so) to improve throughput performance and meet real-time requirements of some mission-critical services. In some cases, the future applications may require RTT values even lower than 1 ms.
- *Download time*, measuring the response time to end-user requests, mostly the web server time to display a web page to the end-user.
- *Video delay*, a live streaming indicator that measures the time, amounting in seconds or several ten seconds, between the instant the camera captures a video frame, and the time the end-user device displays it on the screen.

Layer 4 of the IP protocol stack (end-to-end applications transport layer) mainly manages the applications KPIs. Application KPIs also depend on the network KPIs, those related to the end-to-end transport of the IP packets between the application server and the end-user device. Layers from 1 (physical) to 3 (network) of the IP protocol stack manage the IP packets' transport.

The primary end-to-end network KPIs are:

- *Bit-rate*, a function of the capacity of the network links. Packets flow between the server and the end-user across the network links. The available bit-rate for one application reduces if the number of data streams increases because applications share the bit-rate due to the fairness principle;
- *Packet loss*, the fraction of undelivered packets over the packets sent in a given time frame;
- *Latency*, measured by the round-trip time that is both one of the main network and applications KPIs.

When we consider latency as one KPI at the application layer, it is the value at peak hour one individual end-user experiences. When it is seen as one network layer KPI, we average it over some time and one specific network section to evaluate its proper performance, and possibly correct or repair it.

III. BIT-RATE VS. THROUGHPUT: IMPACT ON INVESTMENTS

As we have seen, the throughput is one application KPI, and the bit-rate is not while it is a network KPI. The bit-rate cannot be an application KPI as it is, by definition, evaluated at Layer 1 (or, for some limited purposes, at Layer 3). Therefore, it does not directly affect the customer experience at the application layer.

*A. Throughput, Latency, and Services*

When a network is narrowband (or broadband), as it was the case for the "old" internet of messages, bit-rate and throughput were numerically equivalent. In practice, it was not a mistake to interchange them. Today, with the "new" internet of contents, and even more in the future with the internet of fast reacting control systems (tactile internet), the situation is much different, and we must take apart the two concepts, as throughput is generally much lower than bit-rate. Throughput is what a customer is interested in, although we all speak of bandwidth (or, even more generically, speed). So for the "old" internet (numerically) *bandwidth=bit-rate*, while for the "new" internet *bandwidth=throughput*.

In other words, bit-rate is the "speed" in the communications channel. Throughput is the "speed" at the level of application, and it is the speed eyeballs experience.

Based on its definition, throughput is easily measured at the application layer. According to [2] it can be "*estimated at the receiver by dividing the number of received bytes by the data transfer completion time. It should not be computed at the sender, because the sender might underestimate the completion time. In particular, the measure of the completion time at the sender may not take into account the time required for emptying the pipe and delivering all the data on the fly*". However, normally the Telco operators do not have access to the application layer. For a Telco, how to qualify network quality based on end-users throughput is still an open issue.

The numerical difference with bit-rate is due to the need to ensure data stream reliability, one task of Layer 4 (transport layer). The network may lose packets during transport due to network congestion and bit errors. The transport layer protocol checks the data integrity to ensure reliable transport service. Then, it sends one ACK (or, equivalently, NACK) message to the sender to retransmit lost or corrupted data, to restore the correct data stream at the expense of added latency. When the bit-rate is sufficiently low in one network, this mechanism does not affect transferring data to the application layer. However, in UBB communications throughput may be limited as a function of a 'distance' metric.

For TCP (Transfer Control Protocol), the most diffused Layer 4 protocol with more than 95% of today's use, the distance metric is $RTT \cdot \sqrt{PLR}$, where $RTT$ is the round trip time (i.e., the measure of latency) and $PLR$ is the packet-loss rate. The model used for *Throughput* estimate, generally known as the Mathis formula, [3] applies whenever TCP's performance is determined solely by the Congestion Avoidance (CA) algorithm:

$$Th(CA) = \frac{c \cdot MSS}{RTT \cdot \sqrt{PLR}} \quad (1)$$

where $c$ is a constant (typical values 0.9÷1.2), and *MSS* is the maximum segment size. Generally, the maximum possible *MSS* is used to minimize header-related transmission overhead and packet-related processing overhead without fragmentation (*MSS* = 1450 Byte). When the distance metric is 'large', *Throughput* is provided by (1), otherwise it is the bit-rate, *BR*, in the channel, i.e., we have:

$$Throughput = \min\{Th(CA), BR\} \quad (2)$$

Figure 1 is based on a well-known representation of services to be delivered with UBB networks [4]. It expresses the need for speed (that according to our clarification is the throughput or, as generally called, the bandwidth) and latency for the use of services by a single user. Each "dot" (i.e., service) represents the (minimum) user requirement and does not consider any network limitations. Therefore, we modified it to show the effect of the dependence of bandwidth (i.e., throughput) on latency for a fixed *PLR*. According to eq. (2), this evidences two different regions: the bit-rate limited region (*Throughput* = *BR*) and the latency-limited region (*Throughput* < *BR*) separated by a boundary pictorially shown with a dashed line.

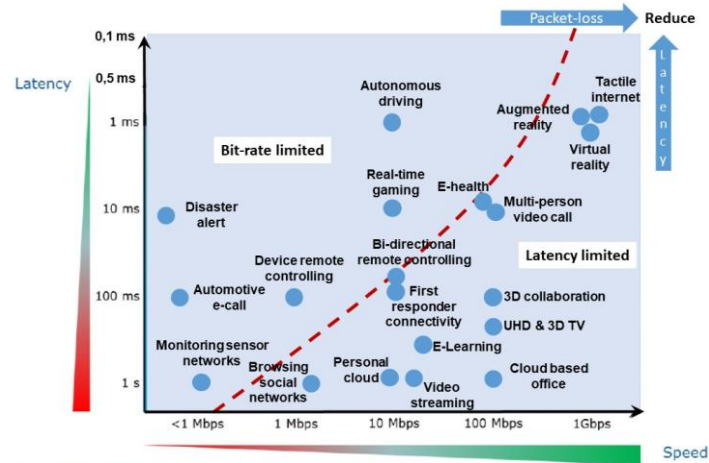

*Figure 1: Applications' bandwidth and latency requirements.*

In the bit-rate limited region, the achievable speed is maximum, and bandwidth and latency are independent. As an example, real-time gaming at 10 Mbps throughput, can be provided as long as the latency is lower than about 100 ms. Certainly, this region is where we like our "dots" may lay. Conversely, as an example, 3D collaboration or multi-person video call, at about 100 Mbps, need a latency lower that 10 ms. Thus, the dots located in the latency-limited region experience degradation of throughput and, for the given values of *RTT* and *PLR*, the network cannot provide the *bandwidth* quoted in abscissa, which is the minimum value for the requested QoE. In this region, throughput depends on latency, bit-rate can be mostly wasted, and the customer cannot perceive the necessary QoE. This is a consequence of "speed" being dependent on "latency" as per eq. (1).

Consequently, for services located in this region, the Telco operator's efforts to increase the bit-rate have no practical effects, while network costs (TCO) may dramatically increase. For example, with *RTT* = 1 ms and *PLR* = 0.1% – both challenging values for today's technology – the achievable throughput cannot be higher than roughly 350-400 Mbps. Therefore, the 1 Gbps speed target is not met (to reach the speed target we need, e.g., *RTT* = 0.5 ms and *PLR* = 0.05%).

Latency reduction is mandatory to achieve the targets of the European Gigabit Society. In practice, in our networks often latency reduction is more urgent than physical layer bit-rate increase. It implies moving servers towards the end-users edge (access network), well inside the Telco operator's network. This change in network architecture can also bring a beneficial effect on *PLR* reduction. Reducing latency (and possibly *PLR*, too) is the objective of Edge-Cloud Computing (ECC). In an UBB network, increasing the bit-rate without revisiting the architecture with the inclusion of ECC is a waste of Telco's investments and cannot provide customers with a sensible increase in technical QoE.

*B. Where is the bottleneck for speed?*

Above we saw that not always bit-rate is the limit for application QoE, as the congestion avoidance algorithm can limit speed at the application layer. Therefore, for the so-called "bandwidth-hungry" services – that are indeed "latency limited" services – it may happen that a bit-rate increase is ineffective if latency is not low enough (for a given packet loss rate).

However, we must take into account one more effect: available user bit-rate is not constant along the path from the source to the end-user, especially in the peak hour. In the residential user access network, in particular in the backhaul link between the first concentration node (such as a DSLAM or OLT) and the first Ethernet switch, the communication capacity is used by simultaneous data streams referred to the users connected to the first concentration node. In addition, in the core network the communication resources are shared in the path between the IP edge, located at the border with the big internet, and the access network backhauling section, between the end-users connected to many concentration nodes. In general we can define the simultaneity coefficient, $s_c$ (%), as the percentage of users served by a network node (in other words, all those connected to that node including those that are not communicating, $N$) that simultaneously transfer data packets at a certain time of the day: so, $s_c = N_c/N$, where $N_c$ denotes the simultaneously communicating users. Neglecting data overheads and assuming respect of the fairness principle, the available bit-rate per user is

$BR = BR_a/(s_c*N)$, where $BR_a$ is the channel capacity (i.e., the overall available bit-rate) of the node to which the users are connected. Such coefficient is different in the backhauling and in the core network sections. In general, network providers assume a certain maximum number of users connected that transfer data at the same time, that is $N_c$. According to this assumption, they reserve a certain amount of bandwidth per user. As an example, if we have a backhauling link at 1 Gbps to which are connected 1000 customers and we assume a maximum simultaneity coefficient of 10% (that is not more than 100 customers that transfer data at the same time), each user will actually use a bandwidth of 10 Mbps in the peak time (1 Gbps/100). In the case of ADSL connection this value can be, in the average, reasonable in the downstream. Of course, if the assumption is not correct, that is the simultaneously coefficient is higher, or if the operator installs a link with lower channel capacity than 1 Gbps, we have a bottleneck. In other words, the available capacity per user is much lower than the physical layer access speed. Of course, in the case of a VDSL access technology the available bit rate per use should be, as an example, 50 Mbps in the peak time. Thus, the network provider should install a backhaul link of at least 5 Gbps. The same concept applies in the core with the difference that in this case the simultaneity coefficient can be lower than in the access network.

Therefore, even if the access physical layer bit-rate is very high (e.g., 1 Gbps), the bottleneck for bandwidth can be in the backhauling or in the core network at the peak-hour. Increasing this access physical layer capacity, installing optical fiber, without upgrading routers speed in packet/s and communication link capacity, can be expensive and ineffective.

*C. One instructive case: the WSJ experiment*

The Wall Street Journal published the results of one research conducted with researchers at Princeton University and the University of Chicago. WSJ conducted on-field tests on internet use in 2019 over months, registering the experience of several ten journalists across the United States. The journalists used only a fraction of their available bandwidth to watch streaming services, including Netflix, Amazon Prime Video, and YouTube, even simultaneously. According to WSJ, "*Quality didn't improve much with higher speeds. Picture clarity was about the same. Videos didn't launch quicker.*" [5]

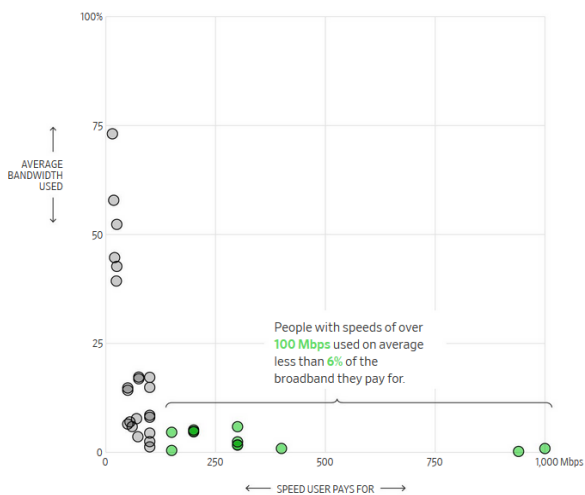

*Figure 2: Share of capacity used during the WSJ test campaign of year 2019 (Source: NYT, 2019).*

In one example case reported in the newspaper's article, a journalist having a speed package of 150 Mbps evidenced that the median usage over 35 viewing minutes was 6.9 Mbps (5% of total capacity). For the portion of time when seven simultaneous streams were going on, the overall average speed was 8.1 Mbps. Figure 2 depicts the statistical evidence of the global situation. Nick Feamster, University of Chicago network-performance expert and one member of the research team, summarizes the main WSJ conclusion: "*For many people they are not going to see huge differences between 50 Mbps, 100 Mbps and a gigabit per second*."

How can we explain what happened in the experiment, and what happens to most of us in the day-by-day experience? Of course, there should be one bottleneck for speed. End-to-end throughput can be limited either in the core network or in the access network when overall available bit-rate is not enough. In the WSJ experiment, the bit-rate was limited inside the core or backhaul network, as the access network bit-rate was enough to serve the multiple simultaneously active data streams.

Very often, the bit-rate limitation happens outside the access network, as was for the WSJ experiment. As another example, Figure 3 shows, in a situation of no packet clipping occurring in the peak time, the measured average peak-hour bit-rate per line connected, that is $BR/N$, to concentration nodes (DSLAM, OLT) in Italy as measured in 2015-2017 and published by the Italian regulator AGCOM [6]. As an example, in the case of a simultaneity coefficient of 10%, the available bit rate per user can be obtained by multiplying the graph value by $N_c$. If $N$=500, that can be a typical value for DSLAM or OLT, $N_c$=50 and the available bit rate per user transferring data is 50 Mbps for UBB in December 2017. The picture also shows that the amount of bandwidth that has to be made available to users increases, in order to avoid packet clipping in the peak time, by 25-30% per year. Thus, using the same parameters of the previous example, the available bandwidth per user transferring data ($N_c$) in 2020 should be, for UBB access lines, at least 100 Mbps or, in other words, 2 Mbps per user connected ($N$).

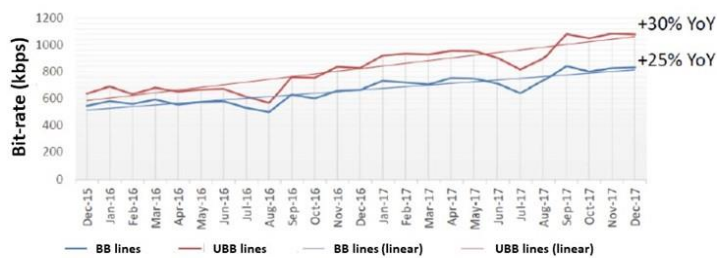

*Figure 3: Average peak bandwidth/total lines in Italian fixed networks measured over two years (Source: AGCOM, 2018)*

IV. THE ROLE OF EDGE-CLOUD COMPUTING

As discussed above, some important UBB services we aim to provide in the future (the year 2025 and beyond) lay in the "latency-limited region" of the service plane. In Figure 1, critical latency-limited services are those below some 50 ms, which is the typical limitation in today's fixed and mobile networks.

Therefore, increasing bit-rate in the access network without reducing the 'distance' metric is useless and, even, a waste of Telco's investments. Today, we should better aim at improving applications KPIs without critically increasing the network cost, which depends on the peak hour load. Due to internet traffic exponential growth YoY, and to service requirements becoming more and more stringent, this situation tends to become progressively more critical. Increasing the bit-rate at the physical (or network) layer does not help much.

According to Eqs. (1) and (2), reducing the 'distance' metric is the correct solution in two respects:
1. Having fixed the desired value for the bit-rate in the access network, eq. (2) shows that *RTT* should reduce so to achieve *Throughput = BR*
2. The most massive traffic bypasses the Telco's core network so that the traffic going through it can enjoy a more substantial fraction of the available bit-rate if the server is located within the access network.

The latter condition is not usual today, as the servers are hosted within the CDNs (Content Delivery Networks), which are located at the Telco's network IP edge or beyond. Therefore, the solution for an ultrafast UBB network is deploying edge-cloud computing platforms to achieve lower round-trip time (and reduced *PLR*). By doing so, in Figure 1, services we are interested in move across the boundary between the two regions, out of the penalizing "latency-limited region."

Edge-cloud computing limits the network load above its location. Therefore, the need to invest in the network sections above the access network reduces. If we consider the TCO, we can save on network costs by using ECC as a result of containment of traffic and the needed total peak throughput in the network segment between the ECC and the big internet.

The transparent cache is one main ECC component capable of achieving traffic containment [7]. It locally stores and continuously updates contents, not altering the end-to-end application logic fully transparent to the content provider and the end-user. While already adopted in CDN nodes located at the Telco's core network or farther away, transparent caching is much more effective when brought closer to the end-user.

From a functional point of view, a transparent cache (Figure 4) is a local repository that stores the most popular content (e.g., the most requested high definition hit movies, the currently most clicked video clips or web pages). It supplies such content whenever a nearby end-user asks, after the content provider completes authentication and authorization. By locally storing copies of the most frequently requested content, the transparent cache can increase applications throughput and reduce download time and video delay, thus improving all the KPIs of the technical QoE. However, to work correctly, a transparent cache must be dynamic in selecting the locally most requested content while updating the memory with new content as it detects changes in the user behavior and expectations.

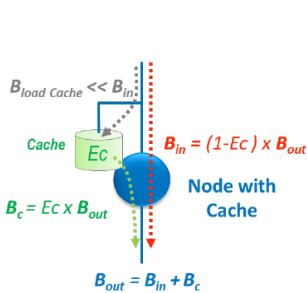

**Legend**. $B_{in}$: input bandwidth (i.e, throughput) offered to the node; $B_{load\ Cache}$: bandwidth used to load contents into the cache; $B_c$: bandwidth provided by the cache; $B_{out}$: total bandwidth towards served clients. We define $E_c$, "*cache efficiency*" (when cache is absent or ineffective, $E_c = 0$ and $B_{in} = B_{out}$). Also, we define "*cache speedup*", ratio of the throughput managed by the node without and with cache, respectively; i.e, $S_u = B_{out}/B_c = 1/E_c$ (if the cache manages half output bandwidth, $S_u = 2$, which means that throughput with transparent cache is two times the one when transparent cache is absent) [1, 7].

*Figure 4: Transparent cache schematic and parameters.*

## V. PARAMETERS FOR THE VHCN DEFINITION

The European Commission promoted the concept of VHCNs with the new European Electronic Communications Code (EECC) to expedite the European Gigabit Society (EGS) objective for the year 2025 [8]. The concept revolves around the development of optical fiber access networks or equivalent. It considers latency (and other technical parameters) *per se*. See Recital (13) in the EECC:

> «*The requirements concerning the capabilities of electronic communications networks are constantly increasing. While in the past the focus was mainly on growing bandwidth available overall and to each individual user, other parameters such as latency, availability and reliability are becoming increasingly important. The current response towards that demand is to bring optical fibre closer and closer to the user, and future 'very high capacity networks' require performance parameters which are equivalent to those that a network based on optical fibre elements at least up to the distribution point at the serving location can deliver. (…)*»

However, when implementing the EECC, it is necessary to consider *the dependence of application throughput on latency*, as discussed in the present paper. The consequence of not considering such dependence is a push toward increasing the bit-rate that is unsuitable or ineffective, when considered alone. Consequently, TCO for Telco operators sharply increases, or concentrates in a short timeframe. Despite that, European citizens gain no or little improvement in QoE. Furthermore, the inappropriate approach in planning investments increases the economic and financial risks for the telecom industry as a whole. It can even backfire against the appreciation of the policy objectives the EC intends to pursue.

The recent (draft) BEREC guidelines at present do not take into account the technical scenario depicted above [9]. The so-called *Criterion 3* and *Criterion 4* of the (draft) VHC guidelines provide "speed" specification in terms of "data-rate" instead of "bandwidth", i.e. throughput. There is time, however, and according to on field evidences, to reconsider the definition of some crucial parameters to improve quality offered to customers at an affordable cost.

The starting point is the correct interpretation of EECC's Recital (13), recognizing the central role of latency (i.e., *RTT*) in the new fast UBB scenario. The EECC does not refer to bit-rate but rather to "bandwidth". The bandwidth concept for today's and future networks cannot be that of "the past", while it must be reconsidered in terms of application throughput, as discussed in the present paper. Moreover, the EECC correctly assumes the point of view of the customer perceived QoE (see Recital (13): «*bandwidth available…to each individual user*»). The customer with the term "bandwidth" is interested in the internet "speed," i.e., the end-to-end throughput.

In fact, as clarified in this paper, the bit-rate in itself is meaningless to the customer, as she only can experience application throughput and never the bit-rate at the physical layer – an engineering parameter fully transparent to her. On the other hand, the Telco operator could decide to provide a specific throughput objective with different combinations of downlink/uplink bit-rate, packet loss ratio, and round-trip packet delay values. While this limitation imposed on the Telco operator does not bring about actual value to the customer, it restricts possible implementation alternatives with an impact on cost and cost intensity and, therefore, on the VHCN coverage extent for a given annual budget. Therefore, a wrong definition of speed in fast UBB networks may damage the EU policy objectives.

Additionally, the bandwidth defined as data-rate along with the assumption that a VHCN needs 1 Gbps (or 150 Mbps for wireless networks) brings about more unintended consequences.

Having regard to the connectivity target for a competitive Digital Single Market (DSM), two EC requirements are important to take into account. First, it is that «*All European households, rural or urban,* [should] *have access to Internet connectivity offering a downlink of at least 100 Mbps, upgradable to Gigabit speed*» [4]. Second is the EECC's Recital (24) that congruently defines VHCNs as «*electronic communications networks which are capable of providing at least 100 Mbps, and which are promptly upgradeable to gigabit speeds*». Thus, when realizing the VHCN, there is an initial target of 100 Mbps mostly relevant in suburban e rural areas. This 100 Mbps target should be a first target to reach. Then, the concept of upgrading the network to 1 Gbps should be correctly interpreted according to the actual UBB network architecture of each.

Thus, the regulator should assess the objectives in terms of throughput having regard to the status of technologies in 2020, and their upgradability. This approach is also essential for FTTH networks of today not to discourage the timely deployment of the new TWDN-PON standard when available on the market.

Table 1 shows the downstream peak-time data-rate for different PON (Passive Optical Network) standards under the common condition of 1:64 splitting factor [10, 11]. To define peak-time under customary conditions, in Table 1 we assumed 99% of the time.

Today's GPON optical networks cannot generally fulfill the 1 Gbps condition on the downstream data-rate when several end-users simultaneously use the network. With a medium filling factor (say, 25%), the data-rate for a GPON is less than 500 Mbps.

## VI. CONCLUSIONS

This paper aimed at highlighting some crucial issues for the VHCNs deployment across Europe, and elsewhere. First, it clarified that for fast UBB networks, the concept of bandwidth is no more equivalent to the customary bit-rate (or, data-rate) at the physical layer, or network layer, but to the concept of throughput at the application layer. Second, it underlined that bandwidth is strictly linked to latency. The most important implication is that some of the most demanding services envisaged for the future (e.g., mobile virtual and augmented reality, tactile internet) cannot be met by simply increasing data-rate. Reducing latency, through ECC, is a mandatory pre-requisite. What discussed above is in agreement with the EECC.

As one main conclusion, each regulator should correctly define down-stream and up-stream bandwidth in terms of throughput rather than bit-rate.

| | $\alpha \rightarrow$<br>Filling $\rightarrow$ | 100% (64 users)<br>MAX | 60% (38 users)<br>LARGE | 25% (16 users)<br>MEDIUM | 8% (5 users)<br>LITTLE |
|---|---|---|---|---|---|
| GPON | $\beta$ | 0% | 6.1% | 56.1% | 97.4% |
| | $B_{99}$ | 147.1 Mbps | 227.3 Mbps | 416.7 Mbps | 1,000 Mbps |
| XGPON | | 63.8% | 97.9% | 100% | 100% |
| | | 588.2 Mbps | 909.1 Mbps | 1,000 Mbps | 1,000 Mbps |
| TWDM-PON<br>(4 colors) | | 100%<br>1,000 Mbps | 100%<br>1,000 Mbps | 100%<br>1,000 Mbps | 100%<br>1,000 Mbps |

Legend:
- Splitting factor 1:64 (equivalently, 2 x 1:8 two-level architecture)
- $B_{max}$ = 1 Gbps (under the assumption that the operator blocks higher bandwidth values)
- $B_{99}$: Bandwidth overcome in 99% of the time (1% unavailability)
- $\beta$ : Prob $\{B = B_{max}\}$
- Activity factor = 15% (if this parameter grows up, $B_{99}$ reduces)
- $\alpha$ : Filling factor
- Case of independent traffic (favorable condition): for correlated traffic, $B_{99}$ is an upper bound.

*Table 1: Downstream data-rate speeds of several optical fiber standards under different operational conditions.*